\documentclass[11pt,twoside
]{article}

\usepackage{baaa2009}
\usepackage{graphicx}
\usepackage{subfigure}
\usepackage{psfrag}
\usepackage{amssymb}
\usepackage[spanish,activeacute,english]{babel}
\usepackage[latin1]{inputenc}
\usepackage[T1]{fontenc} 
\usepackage{ae,aecompl} 
\usepackage{latexsym}
\usepackage{verbatim}
\usepackage{amsmath}
\usepackage{stmaryrd}
\usepackage{amsfonts}
\usepackage{amssymb}
\usepackage{wasysym}
\usepackage[colorlinks=true,dvips]{hyperref}

\begin{document}
\myselectspanish
\vskip 1.0cm
\markboth{J. P. Caso, L. P. Bassino \& A. V. Smith Castelli}%
{Galaxias Enanas Ultra - Compactas (UCD) en el c\'umulo de Antlia}

\pagestyle{myheadings}
\vspace*{0.5cm}
\noindent PRESENTACI\'ON MURAL
\vskip 0.3cm
\title{Galaxias Enanas Ultra - Compactas (UCD) en el c\'umulo de Antlia}


\author{Juan Pablo Caso$^{1}$, 
Lilia P. Bassino$^{1,2}$, Anal\'ia V. Smith Castelli$^{1,2}$}

\affil{%
  (1) Facultad de Ciencias Astron\'omicas y Geof\'isicas, UNLP\\
  (2) Instituto de Astrof\'isica de La Plata (CCT La Plata, CONICET, UNLP); y CONICET \\}

\begin{abstract} 
  We present preliminary results of the search for Ultra-compact 
  dwarf galaxies in the central region of the Antlia cluster. This new 
  kind of stellar system has brightness, mass and size between those 
  observed in globular clusters and early-type dwarf galaxies, but 
  their origin is not well understood yet.
\end{abstract}

\begin{resumen}
  Se presentan resultados preliminares de la b\'usqueda de galaxias enanas
  ultra-compactas (UCD) en la zona central del c\'umulo de Antlia. Este
  nuevo tipo de sistemas estelares posee brillo, masa y tama\~no intermedios
  entre los de c\'umulos globulares y galaxias enanas de tipo temprano, 
  pero a\'un no se comprende exactamente cual es su origen.  
\end{resumen}

\section{Introducci\'on}

\subsection{El c\'umulo de Antlia}

  El c\'umulo de Antlia (l $\approx$ $273$, b $\approx$ $19$, distancia 
 aproximada de $35$\,Mpc) es
  el tercer c\'umulo de galaxias m\'as cercano despu\'es de Virgo y Fornax. 
Posee una riqueza intermedia 
  entre estos dos c\'umulos, pero su densidad en galaxias es a\'un mayor. 

  Antlia posee una estructura compleja, que consiste en dos subgrupos dominados por las 
  galaxias el\'ipticas gigantes NGC\,3258 y NGC\,3268. Los resultados obtenidos hasta la fecha (Smith 
  Castelli 2008 y estudios en rayos X all\'i citados) parecen indicar que las 
galaxias en torno a NGC\,3268 
  constituir\'an un c\'umulo peque\~no, y aquellas situadas en los alrededores 
de NGC\,3258 ser\'an un grupo.

\subsection{Galaxias enanas ultra-compactas (UCD)}
  
  Fueron descubiertas hace una d\'ecada en los alrededores de NGC\,1399, la 
galaxia central del c\'umulo de Fornax (Hilker et al. 1999), siendo 
identificadas luego en otros c\'umulos cercanos. Por 
  tratarse de objetos puntuales o s\'olo marginalmente resueltos a estas 
distancias, en el pasado hab\'ian sido descartadas como objetos de fondo. 
Poseen propiedades intermedias entre los c\'umulos globulares (CGs) 
  y las galaxias enanas de tipo temprano, con brillos intr\'insecos 
-13.5\,<\,Mv\,<\,$-11$\,mag, masas en el rango 
  de $10^6$ a $10^8$\,$M_\odot$ y radios efectivos de $10$ a $100$\,pc; su 
naturaleza no est\'a a\'un claramente establecida (Hilker 2009). 
  
  Entre diversas teor\'ias acerca de su origen, se propone que sean remanentes 
de galaxias 
  enanas nucleadas que han perdido su envoltura por efecto de fuerzas de marea, al ser capturadas por otras galaxias 
  muy masivas (e.g. simulaciones num\'ericas de Bassino et al. 1994), aunque 
no puede descartarse 
  que se trate de CGs {\em genuinos} muy masivos (e.g. G1 en M31 que posee $10^7$\,$M_\odot$, Ma et al. 2009).  
 
\section{Observaci\'on y reducci\'on}
  
  El material observacional consiste en im\'agenes CCD de gran campo ($36'$ x $36'$ y escala de $0.27"$/pixel) de la regi\'on central del c\'umulo de Antlia. Estas fueron obtenidas en el Observatorio Inter-Americano de Cerro Tololo 
  (CTIO, Chile) con la c\'amara MOSAIC. Las observaciones fueron realizadas en los filtros R de Kron-Cousins y C de 
  Washington con un tiempo de exposici\'on de $600$ segundos. 
  
  Para realizar el procesamiento se hizo una primera selecci\'on de objetos 
puntuales con SExtractor sobre la imagen R, 
  previa resta de la luz de las galaxias con un filtro de mediana. Luego se realiz\'o la fotometr\'ia con Daophot/IRAF 
  utilizando una PSF variable sobre todo el campo. Para la calibraci\'on al 
sistema estandar y la correcci\'on por enrojecimiento 
  se utilizaron los valores determinados por Dirsch et al. (2003). Junto a la 
calibraci\'on se convirti\'o el filtro R en T1 del sistema de Washington.

\section{Selecci\'on de la muestra}

  Se acot\'o la selecci\'on a objetos con colores 0.8 < C-T1 < 2.3, rango usual de los CGs (Dirsch et al. 2003, 
  por ej.). Con estos construimos dos muestras; la primera, correspondiente a 
los candidatos a CGs, contiene objetos 
  con magnitudes entre $21$\,$< T1 <$\,$24$. En la segunda, consideramos candidatas a UCDs a aquellos objetos en el rango -13.5\,$< Mv <$\,-11\,mag 
  (Hilker 2009). Asumiendo que el m\'odulo de distancia de Antlia es (m-M) $\approx$ 32.7, y que la relaci\'on entre V y R 
  para galaxias el\'ipticas es V-R = 0.6 (Fukugita et al. 1995), las UCD poseer\'an magnitudes 18.6\,$< T1 <$\,21.1  
  ({\bf Figura 1}). 
  
\begin{figure}[!ht]
  \centering
  \hfill\begin{minipage}[b]{.45\textwidth}
    \centering
    \includegraphics[width=\textwidth]{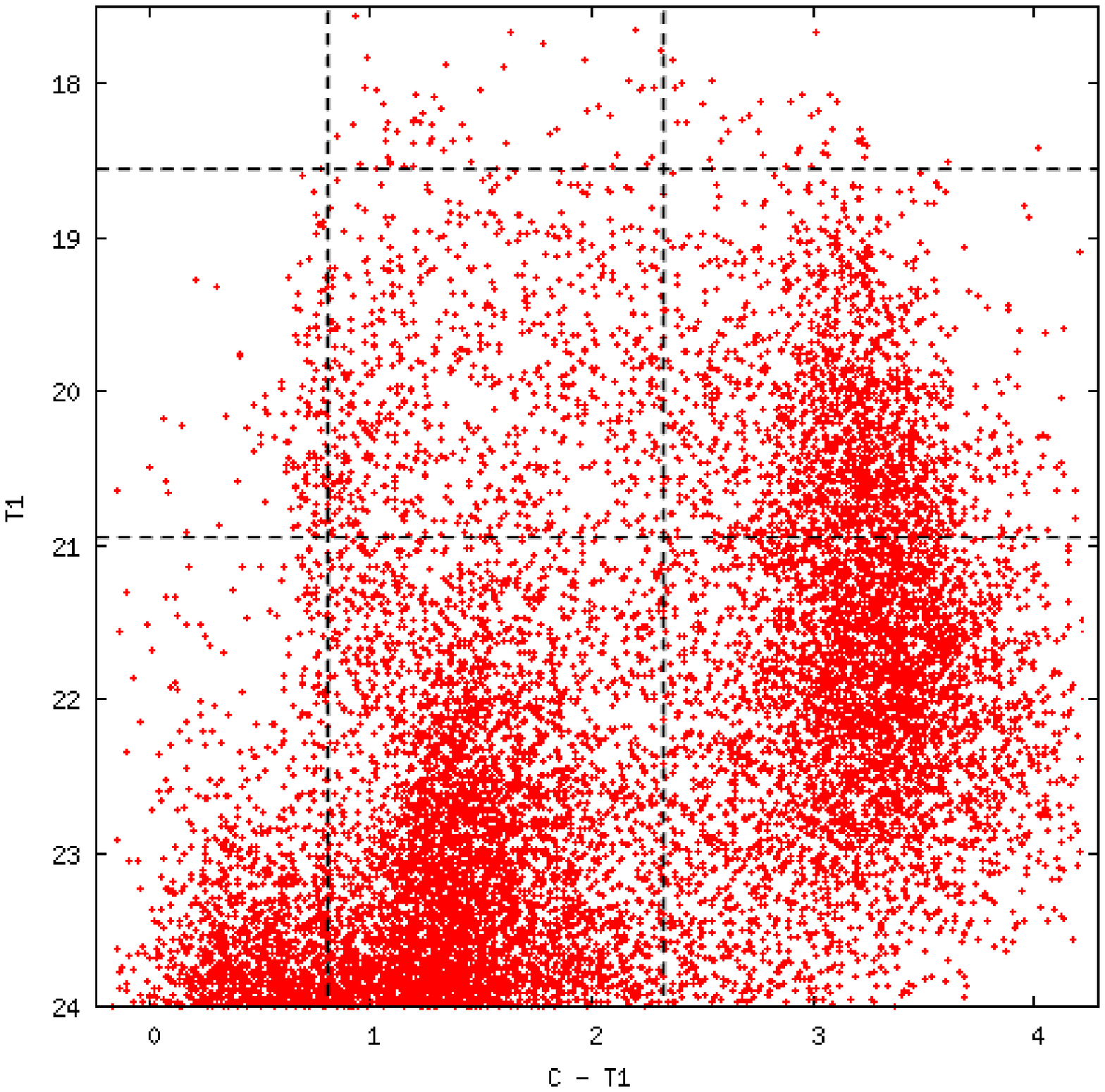}
  \end{minipage}~\hfill%
  \begin{minipage}[b]{.45\textwidth}
    \centering
    \includegraphics[width=\textwidth]{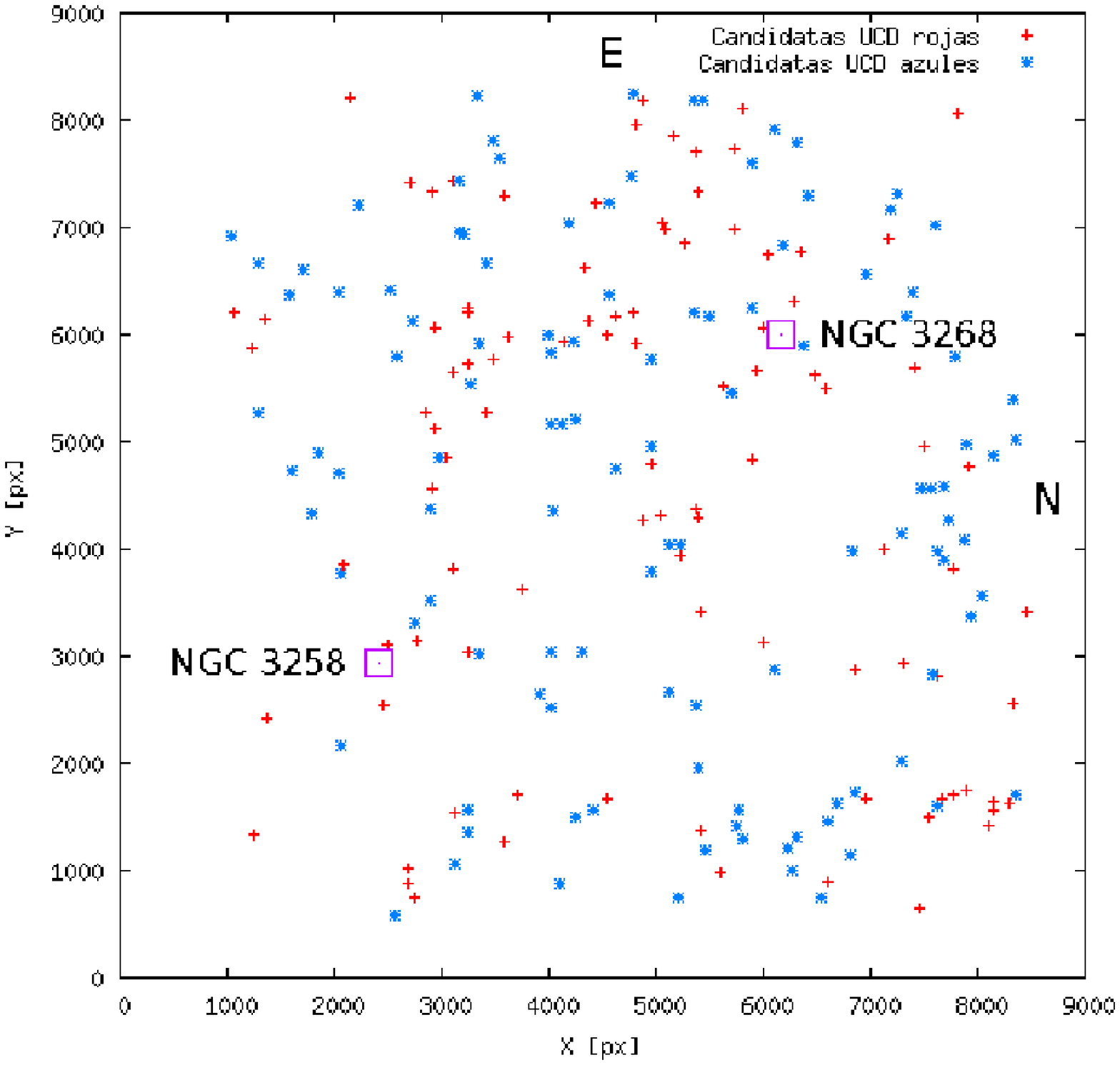}
  \end{minipage}~\hfill~\\[-20pt]
  \hfill\begin{minipage}[t]{.5\textwidth}
    \caption{Diagrama color-magnitud; l\'ineas punteadas 
  delimitan la regi\'on donde se encuentran las candidatas a UCDs}
    \label{fig:ab4}%
  \end{minipage}~\hfill%
  \begin{minipage}[t]{.5\textwidth}
    \caption{Distribuci\'on espacial proyectada de candidatas a UCDs
    en el campo central de Antlia}
    \label{fig:ab5}%
  \end{minipage}\hfill~%
\end{figure}

  La muestra de candidatas a UCDs se compone de $211$ objetos. En la {\bf Figura 2} se muestra su distribuci\'on espacial 
  proyectada, discriminando los candidatos rojos (0.8 < C-T1 < 1.5) y azules 
  (1.5 < C-T1 < 2.3). Las galaxias dominantes est\'an indicadas con sendos cuadrados rosados.

  Para estimar la contaminaci\'on de fondo se consider\'o una regi\'on triangular de 216 minutos cuadrados en la esquina sureste del campo. 
  Debido a la cercan\'ia de esta regi\'on con las zonas de estudio, podr\'iamos estar sobreestimando la contaminaci\'on; debido al reducido 
  tama\~no de la zona y la escasa cantidad de candidatas a UCDs, la estimaci\'on podr\'ia tener errores apreciables.

\section{An\'alisis de la muestra}

\subsection{Distribuci\'on espacial proyectada}

  En la {\bf Figura 2} se observa que el sistema de candidatas en los alrededores de NGC\,$3268$ es el m\'as numeroso. Esto podr\'ia 
  aportar algunos indicios acerca del origen de dichos objetos; recordemos que NGC\,$3268$ ser\'ia, dentro de la estructura del 
  c\'umulo, la galaxia dominante de un c\'umulo peque\~no de galaxias, en tanto NGC\,$3258$ lo ser\'ia de un grupo. Puesto que 
  NGC\,$3258$ posee el sistema de CGs m\'as numeroso, esto podr\'ia indicar una falta de proporci\'on entre los sistemas de CGs y UCDs.

\subsection{Distribuci\'on radial}

  En el caso de las candidatas a UCDs en torno a NGC\,$3258$, al corregir por contaminaci\'on de fondo la poblaci\'on azul se pierde, 
  en tanto la roja se reduce significativamente. Por esto trabajaremos s\'olo con las poblaciones de candidatos a CGs y UCDs cercanas a 
  NGC\,$3268$.

\begin{figure}[!ht]
  \centering
  \hfill\begin{minipage}[b]{.45\textwidth}
    \centering
    \includegraphics[width=\textwidth]{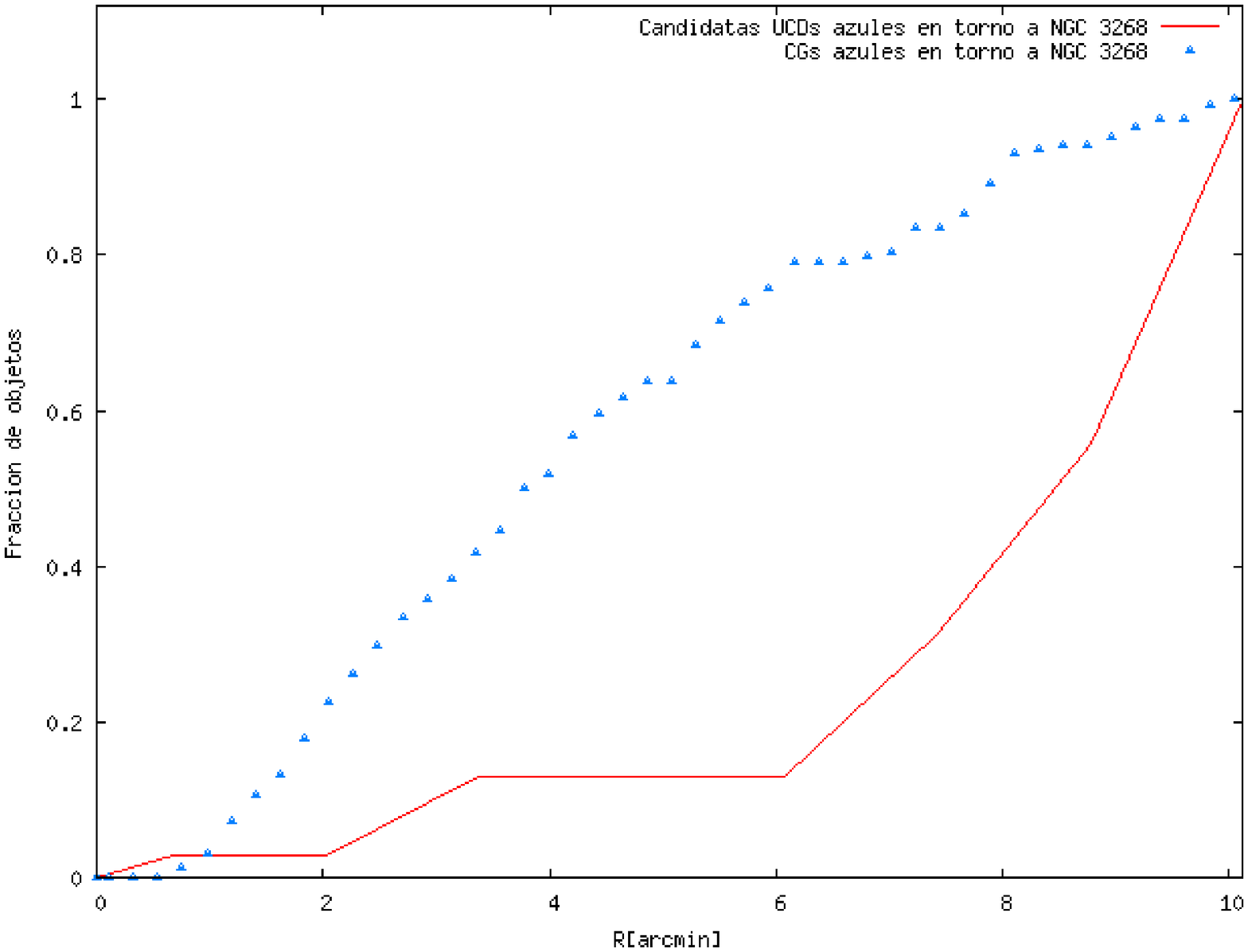}
  \end{minipage}~\hfill%
  \begin{minipage}[b]{.45\textwidth}
    \centering
    \includegraphics[width=\textwidth]{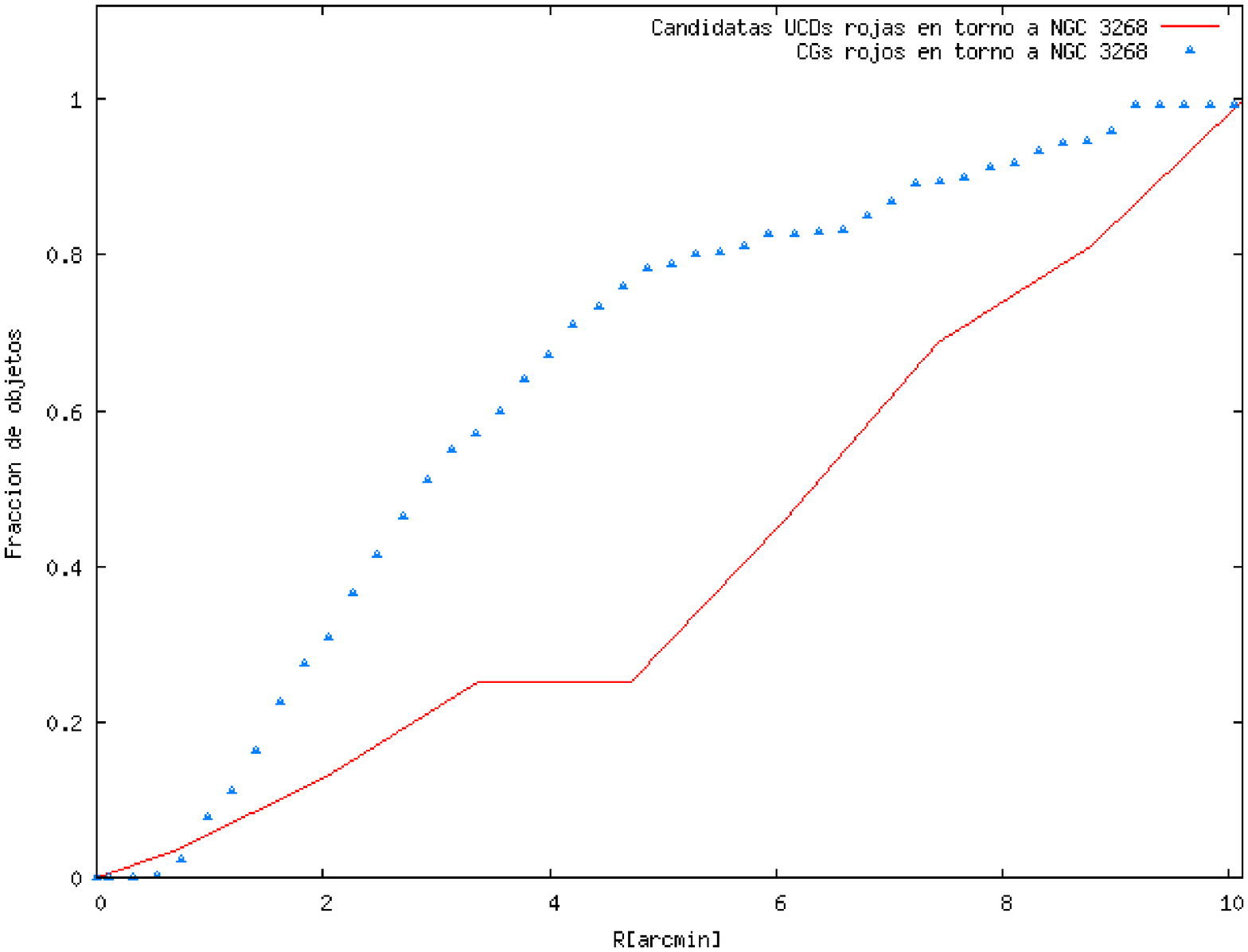}
  \end{minipage}~\hfill~\\[-20pt]
  \hfill\begin{minipage}[t]{.5\textwidth}
    \caption{Distribuci\'on radial proyectada acumulativa de candidatas azules}
    \label{fig:ab4}%
  \end{minipage}~\hfill%
  \begin{minipage}[t]{.5\textwidth}
    \caption{Distribuci\'on radial proyectada acumulativa de candidatas rojas}
    \label{fig:ab5}%
  \end{minipage}\hfill~%
\end{figure}  
   
  En la {\bf Figura 3} se superponen la distribuci\'on radial proyectada de candidatas a UCDs y de candidatos a CGs azules 
  que se hallan a menos de $10'$ de NGC\,$3268$, en tanto la {\bf Figura 4} muestra aquellos objetos rojos.
  En ambos casos las poblaciones de CGs se encuentran m\'as concentradas en torno a NGC\,$3268$ que las candidatas a UCDs de 
  colores equivalentes. Estimamos que tanto la distribuci\'on proyectada de los sistemas de CGs como de las muestras de candidatas a UCDs 
  alcanzan en todos los casos al menos unos $10'$ en torno a NGC\,$3268$.

  Las candidatas a UCDs azules se encuentran menos concentradas que las rojas; apenas el $25$ \% de estos objetos se 
  encuentra a distancias menores que $7'$ de NGC\,$3268$, elev\'andose al $61$ \% cuando estudiamos los objetos rojos. Resultados 
  preliminares indicar\'ian que los objetos m\'as brillantes dentro de la muestra de candidatas a UCDs estar\'an m\'as concentrados 
  en torno a las galaxias dominantes que el resto.

\subsection{Distribuci\'on de color}

  Tras corregir por contaminaci\'on de fondo, la distribuci\'on de color de candidatas a UCDs en torno a NGC\,3268 a\'un presenta objetos
  en todo el rango de colores C-T1 usual de CGs, aunque se observa un claro m\'aximo en el intervalo de color m\'as rojo.
  Parecer\'ia existir una distribuci\'on bimodal, pero su m\'aximo en el rojo no concuerda con el que hemos encontrado para el sistema
  de CGs de esta misma galaxia. Calculamos la magnitud promedio de los objetos situados en cada intervalo, y obtuvimos una 
  tendencia a que objetos m\'as rojos sean m\'as brillantes, resultando una magnitud promedio de 20.1 para el intervalo 0.8\,$< C-T1 <$\,1.05 
  y de 19.5 para el intervalo 2.05\,$< C-T1 <$\,2.30

\begin{figure}[!ht]
  \centering
  \includegraphics[width=.5\textwidth]{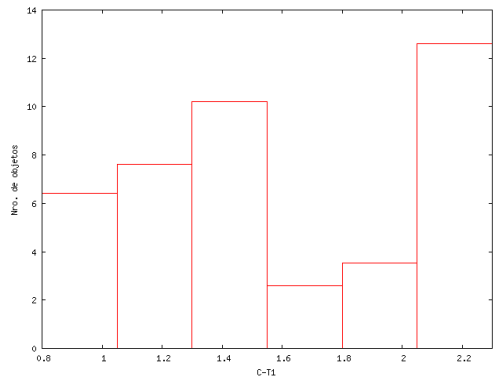}~\hfill 
  \caption{Distribuci\'on de color para candidatas a UCDs en torno a NGC\,3268}
  \label{fig:ab1}
\end{figure}

\begin{referencias}

\reference Bassino L. P., Muzzio J. C., \& Rabolli M. 1994, \apj, 431, 634
\reference Dirsch B., Richtler T., \& Bassino L.P. 2003, A\&A 408, 929
\reference Fukugita M., Shimasaku K. \& Ichikawa T., 1995, PASP, 106, 945
\reference Hilker M., Infante L., Vieira G. et al. 1999, A\&AS, 134, 75  
\reference Hilker M. 2009, ''Reviews in Modern Astronomy'' of the Astron. Gesell., S.Roeser (ed), vol.21, Wiley-VCH,
en prensa (arXiv 0906.0776) 
\reference Ma J., de Grijs R., \& Fan Z. et al. 2009, RAA, 9, 641   
\reference Smith Castelli A.V. 2008, Tesis Doctoral, FCAG (UNLP)                                              
\end{referencias}

\end{document}